\documentclass[12pt]{article}
\usepackage{graphicx, epsfig}
\usepackage[english]{babel}
\begin{document}
\title{Vacuum Polarization in the Presence of Magnetic Flux at Finite Temperature in the Cosmic String Background}
\author{J. Spinelly$^{1}$ {\thanks{E-mail: jspinelly@uepb.edu.br}}  
and E. R. Bezerra de Mello$^{2}$ \thanks{E-mail: emello@fisica.ufpb.br}\\
1.Departamento de F\'{\i}sica-CCT\\
Universidade Estadual da Para\'{\i}ba\\
Juv\^encio Arruda S/N, C. Grande, PB\\
2.Departamento de F\'{\i}sica-CCEN\\
Universidade Federal da Para\'{\i}ba\\
58.059-970, J. Pessoa, PB\\
C. Postal 5.008\\
Brazil}

\maketitle

\abstract{In this paper we analyse the vacuum polarization effect associated with the charged massless scalar field, in the presence of magnetic flux at finite temperature, in the cosmic string background. We consider a spacetime of an idealized cosmic string which presents a magnetic field confined in a cylindrical tube of finite radius. Two different situations are taken into account in our analysis: (i) a homogeneous field inside the tube and (ii) a magnetic field proportional to $1/r$. In these two cases, the axis of the infinitely long tube of radius $R$ coincides with the cosmic string. Specifically, we calculate the effects produced by the temperature in the renormalized vacuum expectation value of the square of the charged massless scalar field, $\langle\hat{\phi}^{\ast}(x)\hat{\phi}(x)\rangle$. Therefore, in order to realize these analysis, we calculate the Euclidean Green function associated with this field in this background.}

\newpage
\section{Introduction}

Our objective in this paper is to investigate the effect produced by the geometry of the spacetime and a magnetic flux on the vacuum polarization effects associated with a charged massless scalar field, at finite temperature.

Apart the thermal effect, the complete analysis about the behavior of a quantum charged field in the neighborhood of an Abelian vortice, must take into account the influence of the geometry and the magnetic field produced by this object. In this way, two distinct cases can be analysed: $i)$ In the first one, the vortice is considered as being a thin linear topological defect, having a magnetic filed running along it. This case can be treated analytically. $ii)$ In the second case, we consider that the string has the non-zero thickness. Analytically this problem becomes intractable. We shall adopt here an intermediate approach. We shall use an approximated model, considering the spacetime produced by the vortice as being of a cosmic string, however having a magnetic field confined into an infinitely long tube of radius $R$ around it. In this way, some improvement is introduced when compared with the ideal case. Because the configuration of vortice magnetic field can not expressed analytically, in attempt to describe the real situation, we shall admit specifics spacial behaviors for the magnetic field inside the tube. The magnetic field along $z$ direction, can be represented by $\vec{H}(r)=H(r)\hat{z}$. Now, assuming that the field is confined in a finite range in the radial coordinate, we are particularly interested in the following two models:    
\begin{eqnarray}
\label{j2}
i)\ \  H(r)&=&\frac{\Phi}{\alpha\pi R^2}\Theta(R-r),\ \ \mbox{homogeneous field inside,} 
\\
\label{j3} 
ii)\ \  H(r)&=&\frac{\Phi}{2\pi\alpha Rr}\Theta(R-r),\ \ \mbox{field proportional to $1/r$ inside.} 
\end{eqnarray}
where $R$ is the radius extent of the tube, $\Theta$ is the Heaviside's function and $\Phi$ is the total flux. The ratio of the flux to the quantum flux $\Phi_{o}$, can be expressed by $\delta=\Phi/\Phi_0=N+\gamma$, where $N$ is the integer part and $0<\gamma<1.$

The analysis of the behavior of the charged massless scalar field in the spacetime of an idealized cosmic string, considering these configurations of magnetics field, has been developed by Spinelly and Bezerra de Mello in \cite{1Spinelly,Spinelly} at zero temperature.

Recently, some authors have investigated the effects of temperature on the vacuum polarization effects in the cosmic string spacetime \cite{Linet, MEX-Linet, MEX}. The standard procedure to introduce temperature effects is by calculating the thermal Euclidean Green function. This can be done for an ultrastatic spacetime\footnote{An ultrastatic spacetime admits a globally defined coordinate system in which the components of the metric tensor are time independent and the conditions $g_{00}=1$ and $g_{0i}=0$ hold} by knowing of the Green function at zero temperature and applying the Schwinger-De Witt prescription \cite{Schwinger}. 

Because we already know the Euclidean Green function at zero temperature \cite{1Spinelly, Spinelly}, we shall adopt this prescription for obtain the thermal Green function. 
 
This paper is organized as follows: In the section $2$ we shall calculate the Euclidean thermal Green function associated with a charged massless scalar field on this system. In the section $3$, using the results obtained in the last section, we shall calculate the thermal renormalized vacuum expectation value of the square of the charged massless scalar field, $\langle\hat{\phi}^{\ast}(x)\hat{\phi}(x)\rangle_{Ren.}$. Finally, we leave for the section $4$ our conclusions.

\section{The Euclidean thermal Green function}
\label{sec2}

The Green function associated with the charged massless scalar field at zero temperature, must obey the following non-homogeneous second-order differential equation
\begin{equation}
\frac{1}{\sqrt{-g}}D_{\mu}\left[\sqrt{-g}g^{\mu\nu} D_{\nu}\right]G(x,x^{'})=-\delta^{(4)}(x-x') \ ,
\label{j5}
\end{equation} 
where $D_{\mu}=\partial_\mu-ieA_\mu$, been $A_{\mu}=(0,0,A(r),0)$.

In order to reproduce the configurations of magnetic fields given by (\ref{j2}) and (\ref{j3}), we must write the third component of the vector potential by
\begin{equation}
A=\frac{\Phi}{2\pi}a(r) \ .
\label{j8} 
\end{equation}

For the two models considered, we can represent the radial function $a(r)$ by:
\begin{equation}
a(r)=f(r)\Theta (R-r)+\Theta (r-R) \ ,
\label{j9}
\end{equation}
with
\begin{eqnarray}
f(r)=\left\{\begin{array}{cc} r^2/R^2,&\mbox{for model ({\it{i}}) and}\\
r/R,&\mbox{for model ({\it{ii}}).}
\end{array}
\right.
\label{j10}
\end{eqnarray}

On the other hand, we assume in our approach that the spacetime produced by the vortice will be the idealized cosmic string one, which, in cylindrical coordinate, reads
\begin{equation}
ds^2=-dt^2+dr^2+\alpha^2 r^2d\theta^2+dz^2 \ , 
\label{jean}
\end{equation}
where the parameter $\alpha$ is smaller than unity and codify the presence of a conical two-surface $\left( r,\theta \right)$.

For the models $1$ and $2$, the solutions of the equation (\ref{j5}), for points outside the magnetic flux, i.e., for $r>R$, are given by \cite{Spinelly}:
\begin{eqnarray}
G^{j}_{T=0}(x,x')&=&\frac{e^{i N\Delta\theta}}{8\pi^2\alpha rr'\sinh u_0}\frac{e^{i\Delta \theta}\sinh(\gamma u_0/\alpha)+\sinh[(1-\gamma)u_0/
\alpha]}{\cosh(u_0/\alpha)-\cos\Delta\theta}
\nonumber\\
&+&\frac{1}{4 \pi^{2} \alpha}\int_0^{\infty} d\omega \omega J_{0}\left(\omega\sqrt{(\Delta 
\tau)^{2}+(\Delta z)^{2}}\right)\times
\nonumber\\
&&\sum_{n=-\infty}^{\infty}e^{i n\Delta\theta}D^j_n(\omega R)K_{|\nu|}(\omega r)K_{|\nu|}(\omega r') , \qquad{j=1,2} \ ,
\label{j19}
\end{eqnarray}
where $\nu =(n-\delta)/\alpha$,
\begin{equation}
\cosh u_{o}=\frac{r^2+{r^{'}}^{2}+(\Delta \tau)^{2}+(\Delta z)^{2}}{2rr^{'}} 
\label{j20}
\end{equation}
and
\begin{equation}
D_n^j(\omega R)=\frac{H_j'(R)I_{|\nu|}(\omega R)-H_j(R)I'_{|\nu|}(\omega R)}
{H_{j}(R)K^{'}_{|\nu|}(\omega R)-H_j'(R)K_{|\nu|}(\omega R)} \ .
\label{j17}
\end{equation} 
In the above equations the functions $H_j(r)$ are given by:
\begin{equation}
H_{1}(r)=\frac{1}{r}M_{\sigma_{1}, \lambda_{1}}\left( \frac{\delta}{\alpha R^2} 
r^2\right),
\label{j15}
\end{equation}
with $\sigma_{1}=(\frac{n}{\alpha}-\frac{\omega^2 R^2\alpha}{2\delta})/2$ and
$\lambda_{1}=n/2\alpha$, and 
\begin{equation}
H_{2}(r)=\frac{1}{\sqrt{r}}M_{\sigma_{2}, \lambda_{2}}\left( \zeta r\right),
\label{j16}
\end{equation}
with $\sigma_{2}=\frac{n\delta}{\alpha}(\delta^2-\omega^2 R^2 \alpha^2)^{-1/2}$, 
$\lambda_{2}=n/\alpha$ and $\zeta=\frac{2}{R\alpha}(\delta^2+\omega^2 R^2 
\alpha^2)^{1/2}$. Moreover, $M_{\sigma, \lambda}$ is the Whittaker function, while $I_{|\nu|}$ e $K_{|\nu|}$ are the modified Bessel functions \cite{tabela}.

Supported by previous analysis, we can determine the thermal Green function, $G_{T}(x,x')$. Following the Schwinger-De Witt prescription. This function is 
\begin{equation}
G_{T}(x,x')=\sum_{l=-\infty}^{\infty}G_{\infty}(x,x'-l\lambda\beta) \ , 
\label{temp}
\end{equation} 
where $\lambda=(1,0,0,0)$ is a Euclidean unitary time-like vector and $\beta=1/k_BT$, being $k_B$ the Boltzmann constant and $T$ the absolute temperature. 

In agreement with the equations (\ref{j19}) and (\ref{temp}), the thermal Green functions associated with the massless scalar field, in the cosmic string spacetime and in the presence of magnetics field described by models $1$ and $2$, are given by: 
\begin{eqnarray}
{G}^{j}_{T}(x,x')&=&\frac{e^{i N\Delta\theta}}{8\pi^2\alpha rr'}
\sum_{l= 0}\frac{e^{i\Delta \theta}\sinh(\gamma u_{l}/\alpha)+\sinh[(1-\gamma)u_{l}/
\alpha]}{\sinh u_{l}\left(\cosh(u_l/\alpha)-\cos\Delta\theta\right)}
\nonumber\\
&+&\frac{1}{4 \pi^{2} \alpha}\sum_{l = 0}\int_0^{\infty} d\omega \omega J_{0}\left(\omega\sqrt{(\Delta 
\tau+l\beta)^{2}+(\Delta z)^{2}}\right)\times
\nonumber\\
&&\sum_{n=-\infty}^{\infty}e^{i n\Delta\theta}D^j_n(\omega R)K_{|\nu|}(\omega r)K_{|\nu|}(\omega r') , \qquad{j=1,2} \ ,
\end{eqnarray}
being
\begin{equation}
\cosh u_{l}=\frac{r^2+{r^{'}}^{2}+(\Delta \tau+l\beta)^{2}+(\Delta z)^{2}}{2rr^{'}}  \ .
\end{equation}

\section{The Computation of $\langle\hat{\phi}^{\ast}(x)\hat{\phi}(x)\rangle_{ Ren.}$ at Non-zero Temperature}
\label{sec3}

The main objective of this paper is to investigate the effects produced by the temperature in the renormalized vacuum expectation value of the square of the charged massless scalar field, $\langle\hat{\phi}^{\ast}(x)\hat{\phi}(x)\rangle$ on this spacetime. In order to do this, we shall take the coincidence limit of the respective Green function. However, this result is divergent. In order to obtain a finite and well defined result, we must apply some renormalization procedure. Here we shall adopt the point-splitting renormalization one. It has been observed that the singular behavior of the Green function has the same structure as given by the Hadamard one, which on the other hand can be written in terms of the square of the geodesic distance between two points. So, here we shall adopt the following prescription: we subtract from the Green function the Hadamard one before applying the coincidence limit as shown below:
\begin{eqnarray}
\langle\hat{\phi}^{\ast}(x)\hat{\phi}(x)\rangle_{T, Ren.}&=&\lim_{x'\rightarrow x}\left[ G^{j}_{T}(x,x')-G_{H}(x,x')\right] \nonumber \\
&=&\langle\hat{\phi}^{\ast}(x)\hat{\phi}(x)\rangle_{T, Reg.}+\langle\hat{\phi}^{\ast}(x)\hat{\phi}(x)\rangle_{T=0}^{C} +\langle\hat{\phi}^{\ast}(x)\hat{\phi}(x)\rangle_{T}^{C} \ .
\label{new}
\end{eqnarray}

The first term on the right side of the above expression, represents the thermal contribution coming from the interaction between charged massless scalar field with a magnetic flux considered as a flux line running along the cosmic string. Fortunately this expression has been obtained by Guimar\~aes in \cite{MEX}, and is given by 
\begin{eqnarray}
\langle\hat{\phi}^{\ast}(x)\hat{\phi}(x)\rangle_{T, Reg.}=\frac{1}{12 \beta}+\frac{1}{16\pi^{2}\alpha r}\int_{0}^{\infty}\frac{\coth \left[ \frac{2\pi}{\beta}r \cosh u/2\right]}{\cosh u/2}F_{\alpha}^{(\gamma)}\left(u,0 \right) \ ,
\label{old}
\end{eqnarray}
where
\begin{eqnarray}
F_{\alpha}^{(\gamma)}\left(u,0 \right)=-2\frac{\sin\left[ \pi\gamma/\alpha\right]\cos\left[ u\left(1-\gamma \right)/\alpha\right]+\sin\left[ u\left(1-\gamma \right)/\alpha\right]\cos\left[ \pi\gamma/\alpha\right]}{\cosh u/\alpha-\cos\pi/\alpha} \ . \nonumber
\end{eqnarray}

The two last terms in (\ref{new}) are corrections on the vacuum polarization due to finite thickness of the radius of the magnetic tube and non-zero temperature. The term $\langle\hat{\phi}^{2}(x)\rangle_{T=0}^{C}$, corresponds to correction only due to finite thickness of the radius of tube, it was given in \cite{1Spinelly,Spinelly}, and reads
\begin{eqnarray}
\langle\hat{\phi}^2(x)\rangle_{T=0}^{C}=-\frac{12\gamma}{\alpha(2\gamma+\alpha)}\left[\frac{w^N_j-\gamma/\alpha z^N_j}{w^N_j+\gamma/\alpha z^N_j}\right]
\left(\frac{R}{r} \right)^{2\gamma/\alpha} \ ,
\label{j33}
\end{eqnarray}
where
\begin{equation}
w^{n}_{1}=\left( \frac{\delta-n}{\alpha}-1\right)M_{\lambda_1,\lambda_2}
(\delta/\alpha)+\left( 1+\frac{2n}{\alpha}\right)M_{\gamma_1,\lambda_1}
(\delta/\alpha) \ ,
\label{j28}
\end{equation}
\begin{equation}
z^{n}_{1}=M_{\lambda_{1},\lambda_{2}}(\delta/\alpha) \ ,
\end{equation}
\begin{equation}
w^{n}_{2}=\left( \frac{\delta-n}{\alpha}-\frac{1}{2}\right)M_{\lambda_2,
\lambda_1}(2\delta/\alpha)+\left(\frac12+\frac{2n}{\alpha}\right)M_{\gamma_2, \lambda_2}(2\delta/\alpha) \ ,
\end{equation}
and
\begin{equation}
z^n_2=M_{\lambda_2,\lambda_1}(2\delta/\alpha) \ ,
\end{equation}
being $\gamma_1=(n+2\alpha)/2\alpha$, $\gamma_2=(n+\alpha)/\alpha$ and
$\nu=\frac{n-N-\gamma}\alpha.$
The term $\langle\hat{\phi}^{\ast}(x)\hat{\phi}(x)\rangle_{T}^{C}$ is the new one, it represents the correction on the expression above due to non-zero temperature. This term can be expressed by
\begin{equation}
\langle\hat{\phi}^{\ast}(x)\hat{\phi}(x)\rangle_{T}^{C}=\frac{1}{2\pi^{2}\alpha r^{2}}S_{j}(r) \ ,
\end{equation}
where
\begin{equation}
S_{j}(r)=\sum_{l=1}^{\infty}\int_0^{\infty} dv v J_{0}\left(v\xi l\right)\sum_{n=-\infty}^{\infty}D^j_n(vR/r)K^{2}_{|\nu|}(v)\  , \qquad{j=1,2} \ ,
\end{equation}
with $\xi=\beta/r$.

Unfortunately we cannot provide a closed expression to $S_j(r)$. The best we can do is to present its behavior as function of $r$ and $T$ numerically. We can shown that the most important contribution in the $n$ sum above comes from $n=N$. On other hand we observe that we can change the exact expressions of the coefficients $D_n^{j}(vR/r)$ by the approximated one. The approximated expressions for the coefficients, which are obtained when we take $R/r \ll 1$, are given by \cite{Spinelly} 
\begin{equation}
D_n^{j}(vR/r)=-\frac{2}{\Gamma(|\nu|+1)\Gamma(|\nu|)}\left(\frac{w^n_i
-|\nu|z^{n}_{i}}{w^{n}_{i}+|\nu|z^{n}_{i}}\right)\left( \frac{v R}{2r}
\right)^{2|\nu|}.
\end{equation}

Consequently the sums $S_{j}(r)$ becomes
\begin{eqnarray}
S_j(r)&=&-\frac{2} {\Gamma(\gamma/\alpha+1)\Gamma(\gamma/\alpha)}\left(\frac{w^n_i
-\gamma/\alpha z^{n}_{i}}{w^{n}_{i}+\gamma/\alpha z^{n}_{i}}\right)\left( \frac{R}{2r}
\right)^{2\gamma/\alpha} \times \nonumber \\
&&\sum_{l=1}^{\infty}\int_0^{\infty} dv v^{1+2\gamma/\alpha} J_{0}\left(v\xi l\right)K^{2}_{|\gamma/\alpha|}(v) , 
\end{eqnarray}

Even considering the approximated expression to the coefficient $D_n^{j}(vR/r)$, we can not obtain an analytical expression for the sum $S_J(r)$. However, through a numerical analysis we can determine dependence of this sums depend with the temperature. Considering $R/r=10^{-3}$, $\gamma=0.2$ and $\alpha=0.99$, we have that the behaviors of the sums $S_j(r)$, for high temperature, are given by figure bellow.  
\begin{figure}[h]
\label{fig10}
\begin{center}
\includegraphics[width=9cm,angle=0]{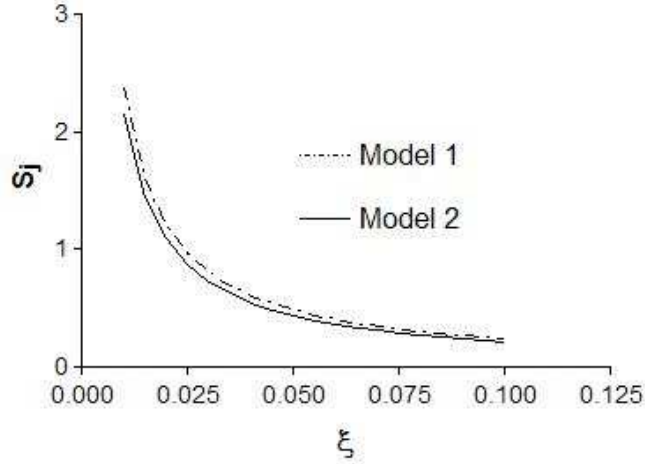}
\caption{Dependence of the sum $S_{j}$ with $\xi=\beta/r$, for the models $1$ and $2$.}
\end{center}
\end{figure}

In this case the dependence of $S_j$ with $\xi$ is 
\begin{eqnarray}
S_j(r)=\frac{a_j}{\xi^{b}} =a_{j}\left(\frac{r}{\beta}\right)^{b} \ , \quad j=1,2 \ .
\label{soma}
\end{eqnarray}
For the first model we found:
\begin{eqnarray}
	a_{1}=0,02465 \pm 0,0003 \ ,
\end{eqnarray}
 while for the second model we found
\begin{eqnarray}
	a_{2}=0,02231\pm 0.0003 \ .
\end{eqnarray}
For both models the value for the exponent $b$ are equal and reads:
\begin{eqnarray}
b=0.99255\pm 0,00325 \ .	
\end{eqnarray}

In the high-temperature regime $(\beta\rightarrow 0)$, the expression (\ref{old}) becomes \cite{MEX}
\begin{equation}
\langle\hat{\phi}^{\ast}(x)\hat{\phi}(x)\rangle_{T, Reg.}\approx \frac{1}{12 \beta^{2}}+\frac{M^{(\gamma)}}{\beta r} \ ,
\end{equation}
where the constant $M^{(\gamma)}$ is defined by
\begin{eqnarray}
M^{(\gamma)}=\frac{1}{16 \pi^{2} \alpha}\int_{0}^{\infty}\frac{F_{\alpha}^{(\gamma)}\left( u,0\right)}{\cosh(u/2)} \nonumber \ . 
\end{eqnarray}
 
Hence, the expression for renormalized vacuum expectation value of the square of the charged massless scalar field, in the high-temperature limit, is given by
\begin{equation}
\langle\hat{\phi}^{\ast}(x)\hat{\phi}(x)\rangle_{T, Ren.}\approx \frac{1}{12 \beta^{2}}+\frac{M^{(\gamma)}}{\beta r} +\frac{a_j}{2\pi^{2}\alpha }\frac{1}{r^{2-b}\beta^{b}} \ ,
\end{equation}

Considering $\gamma=0.2$ and $\alpha=0.99$, we have that $M^{(\gamma)}=-0.0129973$.  

\section{Concluding remarks}
\label{concluding}
In this paper we have analysed the thermal effects on the renormalized vacuum expectation value of a charged massless scalar field, $\langle\hat{\phi}^{\ast}(x)\hat{\phi}(x)\rangle$, in the cosmic string spacetime considering the presence of a magnetic flux of finite radius. Two specific configurations of magnetic flux have been considered. For both we can express the vacuum polarization as
\begin{eqnarray}
\langle\hat{\phi}^{\ast}(x)\hat{\phi}(x)\rangle_{T, Ren.}=\langle\hat{\phi}^{\ast}(x)\hat{\phi}(x)\rangle_{T,Reg.} +\langle\hat{\phi}^{\ast}(x)\hat{\phi}(x)\rangle_{T=0}^{C}+\langle\hat{\phi}^{\ast}(x)\hat{\phi}(x)\rangle_{T}^{C} \ ,
\end{eqnarray}
where the first term on the right hand side represents the thermal contribution coming from the idealized magnetic flux running along the cosmic string. This contribution has been exactly calculated by Guimar\~aes in \cite{MEX}. The second term is the zero temperature contribution on the vacuum polarization due to the non vanishing radius of the magnetic flux tube. This contribution has been analysed in \cite{Spinelly}. The third contribution is the new one. It comes from the combination of the non vanishing radius of the magnetic flux tube and non zero temperature. It goes to zero for $R\to 0$ and for $T\to 0$. Unfortunately this new contribution cannot be expressed in terms of any analytical function. So, for both configuration of magnetic flux, we observed numerically that in the high temperature regime, the thermal contributions due to them present a dependence on the parameter $\xi=\beta/r$ approximately given by $1/\xi$. This means that these contributions are subdominant in the high temperature regime.  
\\       \\

{\bf{Acknowledgment}}
\\       \\
One of us (ERBM) wants to thanks Conselho Nacional de Desenvolvimento Cient\'\i fico e Tecnol\'ogico (CNPq.), and also CNPq./FAPESQ-PB (PRONEX), for partial financial support.


\begin{thebibliography}{99}

\bibitem{1Spinelly} J. Spinelly e E. R. Bezerra de Mello, Int. J. Mod. Phys. {\bf A 17}, 4375 (2002).
\bibitem{Spinelly} J. Spinelly e E. R. Bezerra de Mello, Class. Quantum Grav. {\bf 20}, 873 (2003). 
\bibitem{Linet} B. Linet, Phys. Lett. A {\bf 124}, 240 (1987).
\bibitem{MEX-Linet} M.E.X. Guimar\~aes and B. Linet, Comm. Math. Phys. {\bf 165}, 297 (1994).
\bibitem{MEX} M.E.X. Guimar\~aes, Class. Quantum Grav. {\bf 12}, 1705 (1995).
\bibitem{Schwinger} B. S. De Witt, in {\it{General Relativity: an Einstein Centenary Survey}}, edited by S. W. Hawking end W. Israel (Cambridge University Press, Cambridge, England, 1982); H. W. Braden, Phys. Rev. D {\bf 25}, 1028 (1982)
\bibitem{tabela} I. S. Gradshteyn e I. M Ryzhik, {\it{Table of Integral, Series and Products}} (Academic Press, New York, 1980).
\end{thebibliography}
\end{document}